\documentclass{article}
\usepackage{amsmath}
\usepackage{graphicx} % Add this line
\usepackage{xcolor}
\pagecolor{white}
\color{black}

\definecolor{lightpurple}{rgb}{0.85, 0.85, 1} % Light purple background color
\definecolor{white}{rgb}{1, 1, 1} % White text color

\title{Rugsafe: A multichain protocol for recovering from and defending against Rug Pulls}
\author{Jovonni L. Pharr, Jahanzeb M. Hussain \\ info@rugsafe.io}
\date{v0.0.1 September 1, 2024}

\begin{document}

\maketitle

\begin{abstract}
Rugsafe introduces a comprehensive protocol aimed at mitigating the risks of rug pulls in the cryptocurrency ecosystem. By utilizing cryptographic security measures and economic incentives, the protocol provides a secure multichain system for recovering assets and transforming rugged tokens into opportunities and rewards. Foundational to Rugsafe are specialized vaults where rugged tokens can be securely deposited, and anticoin tokens are issued as receipts. These anticoins are designed to be inversely pegged to the price movement of the underlying rugged token. Users can utilize these anticoins within the ecosystem or choose to burn them, further securing the protocol and earning additional rewards. The supply of the native Rugsafe token is dynamically adjusted based on the volume, value, and activity of rugged tokens, ensuring stability and resilience. By depositing rugged tokens into a vault on several chains, and by burning anticoins, users receive incentives on the RugSafe chain. This protocol's vaults are designed to work in heterogenous blockchain ecosystems, offering a practical and effective solution to one of the most significant challenges in the cryptocurrency market. 
\end{abstract}

%\clearpage

\tableofcontents

%\twocolumn

\section{Overview}

The Rugsafe protocol enables users who hold rugged tokens, denoted as $C_r$, to deposit these tokens into a specialized vault $V_c$ and receive anticoin tokens, $C_a$,  which serve as receipts, \(\mathcal{R} \), and are inversely pegged to the value of the underlying rugged token. The supply of Rugsafe's native token, $R$, is regulated based on the total value of rugged tokens in existence.

%%%%%%%%%%%%%%%%%%%%%%

\section{Definition of a Rug Pull}

A rug pull in the cryptocurrency and decentralized finance (DeFi) space occurs when a token’s value collapses rapidly, leaving users holding worthless or severely devalued tokens \cite{agarwal2023}. While some rug pulls result from deliberate fraud, others may arise from project failures or external market conditions. We define three distinct types of rug pulls, each with unique and quantifiable price dynamics, mirroring the taxonomy proposed in \cite{zhou2022}..

\subsection{Types of Rug Pull}

\subsubsection{Scam Rug Pull}

In a scam rug pull, the creators of the token (the "rug token") deliberately drain liquidity after building trust with users. This is the most direct and malicious form of rug pull, characterized by a sharp, near-instantaneous collapse in token value once liquidity is removed. Mathematically, the token price \( P_{r}(t) \) collapses rapidly after the liquidity event:

\[
P_{r}(t) = P_{r}(0) \cdot e^{-\frac{t}{\tau_{\text{rug}}}}
\]

where \( P_{r}(0) \) represents the initial token price, and \( \tau_{\text{rug}} \) is a small time constant representing how quickly the price collapses after the liquidity is drained. As \( \tau_{\text{rug}} \) becomes smaller, the price falls faster, leading to an almost instantaneous collapse.

\subsubsection{Catastrophic Event Rug Pull}

A catastrophic event rug pull occurs when a legitimate project experiences an unexpected failure, such as a major design flaw or market crash. In contrast to a scam, this type of rug pull results from project failure rather than fraud, leading to an exponential decline in the token’s value:

\[
P_{r}(t) = P_{r}(0) \cdot e^{-\lambda t}
\]

In this equation, \( \lambda \) is a constant representing the rate of decline due to external factors. The token value decays gradually, reflecting the project's collapse over time.

\subsubsection{Legitimate Project Resembling a Rug Pull}

In this scenario, a project remains functional, but due to negative market sentiment or a loss of community trust, the token price behaves similarly to a rug pull. This pattern is especially visible in NFT markets, where floor-prices fade even though projects remain online \cite{huang2023}. Although the project continues to operate, the token price \( P_{r}(t) \) declines hyperbolically, approaching zero as market sentiment erodes:

\[
P_{r}(t) = P_{r}(0) \cdot \frac{1}{1 + \alpha t}
\]

Here, \( \alpha \) represents the rate at which market sentiment deteriorates. This hyperbolic decay models the steady decline of the token’s value, resembling a rug pull, though the project remains active.

\subsection{Mechanism of a Blatant Scam Rug Pull}

\subsubsection{Creation of the Rug Token} 

The token creator mints a large supply of the rug token and provides initial liquidity by pairing the rug token with a more liquid asset, such as ETH or USDC, in a liquidity pool.

\subsubsection{Attracting Liquidity} 

Users are encouraged to deposit liquid tokens into the liquidity pool, drawn by potential rewards or staking opportunities. As the liquidity increases, the token creator builds trust with the community. Tactics often rely on bot networks that artificially inflate volume and social-media buzz \cite{janetzko2023}.

%%%%%%%%%

\subsubsection{Liquidity Drain} 

Once sufficient liquidity is added, the token creator removes liquidity by swapping their large rug token supply for the liquid tokens in the pool. In a simplified model, the liquidity drain is proportional to the creator's rug token holdings.

\[
\Delta L_{\text{pool}} = - L_{\text{liquid}} = \frac{T_{\text{rug}}}{T_{\text{total}}} \cdot L_{\text{pool}}
\]

where \( L_{\text{liquid}} \) is the liquid token withdrawn from the pool, \( T_{\text{rug}} \) is the rug token supply held by the creator, and \( T_{\text{total}} \) is the total rug token supply in circulation.

However, for example, in AMM-based liquidity pools, liquidity is governed by the constant product formula:

\[
x \cdot y = k
\]

where \( x \) and \( y \) represent the amounts of each token in the pool, and \( k \) is a constant. As the creator withdraws liquidity, the price of the rug token \( P_{\text{rug}} \) will increase due to slippage. The larger the withdrawal, the more the rug token price rises, making the liquidity drain less efficient as the creator attempts to withdraw more.

\subsubsection{Outcome for Victims}

After the liquidity is drained, the remaining users hold severely devalued rug tokens, while the creator absconds with the liquid assets. The price of the rug token dramatically collapses, approaching zero but never truly reaching it due to residual speculative trading or illiquidity.

\[
\lim_{t \to t_{\text{rug}}} P_{\text{rug}}(t) \approx \epsilon, \quad \text{where} \, \epsilon \to 0
\]

This equation indicates that after the rug event at time \( t_{\text{rug}} \), the price of the rug token \( P_{\text{rug}}(t) \) approaches a negligible value \( \epsilon \), where \( \epsilon \) is a small positive number close to zero. Though the token may continue to trade, its value is effectively worthless for the remaining holders.

%%%%%%%%%%%
%\subsection{Impact}

%%%%%%%%%%%%%%%%%%%%%%%%%%%%%%%%%%%% ANTICOINS

\section{Vault Creation and Anticoin Issuance}
When a vault $V_c$ is created for a rugged token $C_r$, users can deposit $C_r$ into this vault. Upon deposit, the protocol mints an equal amount of $C_a$ tokens, which are credited to the user's balance. These $C_a$ tokens serve as a new, clean fungible token for the rugged victim, issued 1:1 with the underlying rugged token and are inversely pegged to its value.

\begin{equation}
C_a = C_r
\end{equation}

In addition to the 1:1 $C_a$ tokens, the user may also receive a receipt, \(\mathcal{R} \), in the form of a non-fungible token (NFT) or a refungible token (RFT). The 1:1 $C_a$ token acts as a standard fungible token, fully transferable and tradable. However, the receipt could also be issued as an NFT, representing a unique claim on the deposited rugged tokens. Alternatively, the receipt can be an RFT, where an NFT representing the deposit is held by a contract. The contract then distributes shares of the NFT by issuing a fungible token that serves as representative shares of the underlying NFT.

The type of asset used as the receipt, whether fungible, non-fungible, or refungible, is parameterized at the time of vault creation, with the relationship expressed as:

\[
\mathcal{R} = f(C_r, \tau)
\]

where $\tau$ determines whether the receipt is a fungible token, an NFT, or an RFT. This provides flexibility depending on the underlying asset type. The vaults are held in a central vault registry per blockchain ecosystem, and ultimately represented on the Rugsafe blockchain, ensuring that all interactions with rugged tokens are tracked and managed within the multichain ecosystem. Regardless of the type of receipt issued, depositors always receive the 1:1 $C_a$ tokens as a clean representation of their original rugged tokens.

%%%%%%%%%%%%%%%%%%%%%%%%%%%%% ANTICOIN ROLE

\subsection{Anticoin Role and Market Interaction}

At the time of vault creation, the protocol mints $C_a$ tokens in a 1:1 ratio with the deposited rugged tokens $C_r$. These $C_a$ tokens serve not only as receipts for the deposited tokens but also as inversely, logarithmically pegged assets. This inverse logarithmic peg allows users to benefit from a decline in the value of $C_r$, effectively creating a hedge against further depreciation of rugged tokens.

\begin{figure}[h]
\centering
\includegraphics[width=\textwidth]{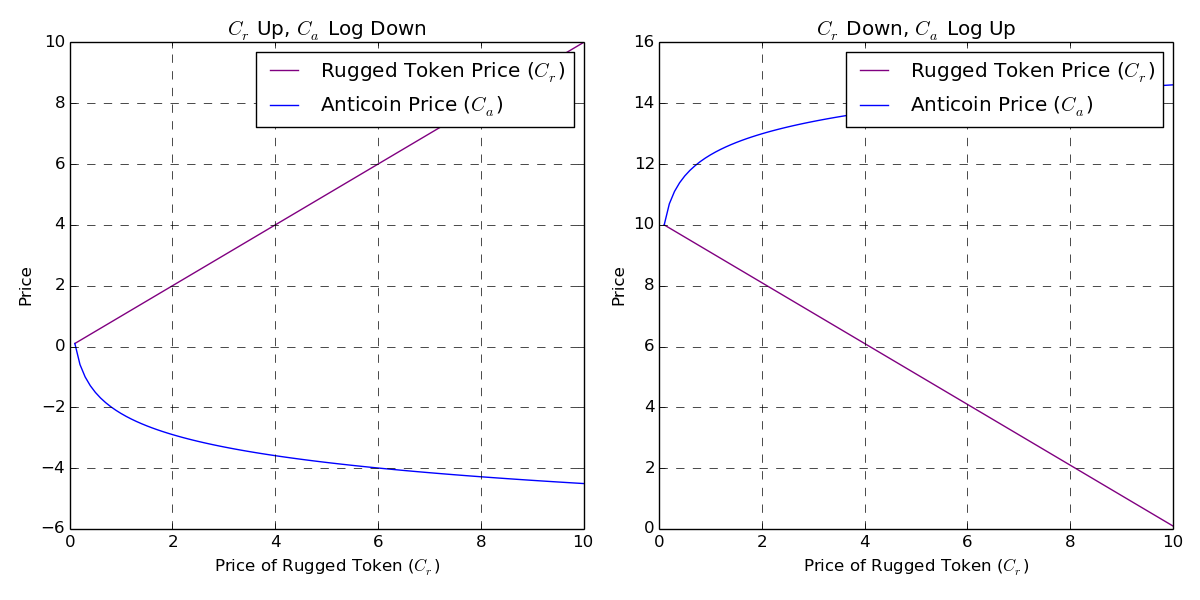}
\caption{The plotted relationship between the price of a rugged token ($C_r$) and the price of its anticoin ($C_a$) under an inverse logarithmic model. In the left plot, as $C_r$ increases linearly, the price of $C_a$ decreases logarithmically, demonstrating a diminishing response in $C_a$ as $C_r$ grows. In the right plot, as $C_r$ decreases, the price of $C_a$ rises logarithmically, tapering off as $C_r$ approaches zero. This inverse peg mechanism ensures that $C_a$ responds proportionally to changes in $C_r$, maintaining stability and mitigating volatility.}
\label{fig:anticoins_plot}
\end{figure}

Unlike traditional stablecoins, the value of $C_a$ tokens increases as the value of $C_r$ decreases, following an inverse, logarithmic price relationship. This dynamic creates a unique market interaction where $C_a$ holders are incentivized to retain their tokens when the value of $C_r$ declines.

The inverse logarithmic peg mechanism, as depicted in Figure \ref{fig:anticoins_plot}, ensures that $C_a$ tokens adjust logarithmically based on the real-time market movements of the underlying $C_r$. Unlike bounded inverse assets, the $C_a$ tokens are not constrained by any upper or lower price limits, allowing for unrestricted value increases as $C_r$ approaches zero.

This design ensures that $C_a$ tokens serve dual purposes within the protocol. They serve as a receipt for deposited rugged tokens and as an inversely, logarithmically pegged asset that can be freely traded or held. The valuation of $C_a$ is driven by market forces, allowing the broader ecosystem to determine its worth relative to other assets based on supply, demand, and the inverse relationship with the rugged token $C_r$.

%%%%%%%%%% bounding
\subsection{Inverse Logarithmic Pegging as Soft Theoretical Bounds}

Linear inverse pegs in previous systems required artificial bounds on asset prices. Once these bounds were reached, the asset froze, and a new version was created. These restrictions were designed to prevent extreme price volatility but imposed artificial constraints that limited flexibility and created friction through asset freezes, and resets.

In contrast, the inverse logarithmic peg for $C_a$ naturally establishes soft theoretical bounds without requiring such freezes. As the price of $C_r$ moves to either extreme, the logarithmic peg ensures the price of $C_a$ tapers off, maintaining stability. This tapering effect acts as a soft boundary, ensuring proportional price responses without hard limits or resets.

\[
C_{a}(t) = \log\left(\frac{C_{r}(0)}{C_{r}(t)}\right)
\]

Where \( C_{a}(t) \) is the value of the anticoin at time \( t \), \( C_{r}(0) \) is the rugged token's price at vault creation, and \( C_{r}(t) \) is its real-time price. This relationship ensures that all $C_a$ holders share a uniform reference point.

This approach provides more organic price regulation, allowing the market to drive $C_a$’s value while avoiding the volatility issues seen in bounded systems. The logarithmic peg offers a flexible, stable framework that adapts naturally to market conditions.

%%%%%%%%%%%%%%%%%%%%%%%%%%%%%%%
%%%%%%%%%%%%%%%%%%%%%%%
%%%%%%%%%%%%%%%%%

%%%%%%%%%%%%%%%%%%%%%%%%%%%% SUPPLY

\section{Supply Regulation Mechanism}
The protocol regulates the supply of the native Rugsafe token $R$ based on the total value of all rugged tokens $C_r$ in existence. The supply of the Rugsafe token is inversely related to the sum of the values of these rugged tokens, adjusted logarithmically:

\begin{equation}
R_\text{supply}  \propto \frac{1}{\log\left(\sum C_r\right)}
\end{equation}

This mechanism ensures that as more tokens are rugged, the supply of $R$ becomes increasingly scarce, potentially enhancing its value and stability within the ecosystem.

\begin{figure}[h]
\centering
\includegraphics[width=\textwidth]{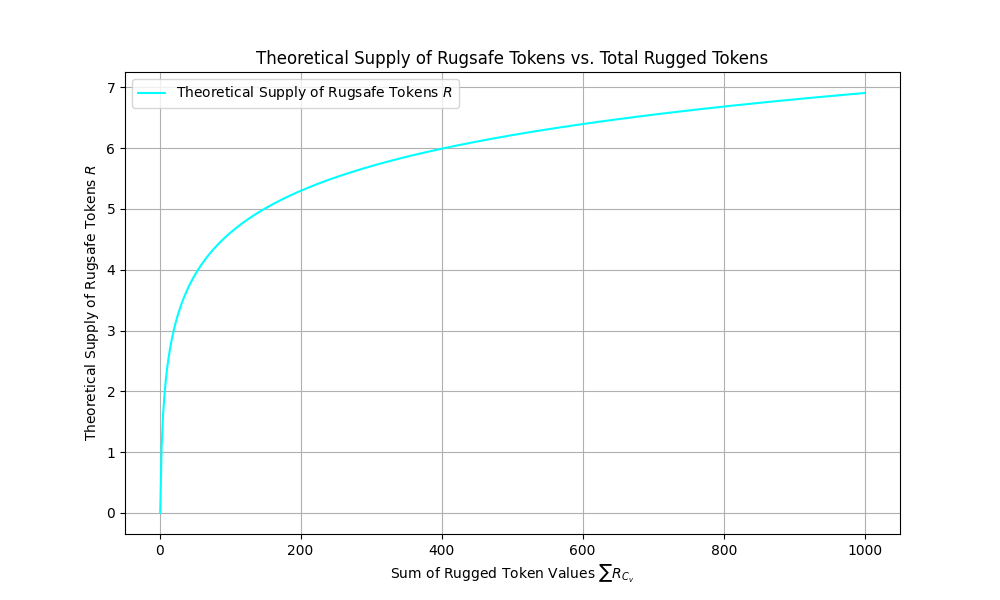}
\caption{The logarithmic relationship between the theoretical supply of Rugsafe tokens $R$ and the sum of rugged token values $\sum C_r$. This relationship underscores how the supply of $R$ is regulated based on the value of rugged tokens in existence, ensuring stability and scarcity.}
\label{fig:minted_anticoins}
\end{figure}

In addition to this scarcity mechanism, the native chain introduces emissions of Rugsafe tokens $R$ per block produced, as it operates on the Cosmos SDK framework. This implies that the total supply of $R$ is influenced by two opposing forces: the burning of tokens $R_\beta$ and the minting of new tokens $R_\texttt{mint}$ through block emissions.

The rate of token burning is governed by the inverse relationship with the total value of rugged tokens $C_r$ in existence:

\[
R_{\beta} \propto \frac{1}{\log\left(\sum R_{C_v}\right)}
\]

Simultaneously, the protocol continuously mints new tokens at a rate proportional to the number of blocks produced:

\[
R_{\texttt{mint}} = \epsilon \times b
\]

where $\text{emission\_rate}$ is a parameter that determines the number of Rugsafe tokens emitted per block.

The net change in the supply of Rugsafe tokens $R_\text{net}$ over time is the difference between the tokens minted and burned:

\[
R_{\text{net}} = R_{\epsilon} - R_{\beta}
\]

Both the burn rate $R_{\beta}$ and the emission rate per block $R_\epsilon$ are parameterized and can be adjusted through community governance:

\[
R_\beta = f\left(\frac{1}{\log\left(\sum R_{C_v}\right)}, \beta_{\texttt{burn}}\right)
\]

and to calculate the amount minted, we can take the emission rate per block, multiplied by the amount of blocks considered,

\[
R_{\beta} = g(\epsilon, \text{blocks})
\]

This flexibility allows the community to respond to changing market conditions and fine-tune the protocol to achieve the desired balance between token scarcity and availability. By giving the community control over these parameters, the protocol ensures that it can adapt over time, maintaining the health and stability of the Rugsafe ecosystem.

%%%%%%%%%%%%%%%%%%%%%%%%%%%%%%%

\section{User Interaction and Ecosystem Signaling}
When users deposit rugged tokens $C_r$ into a vault $V_c$, they receive $C_a$ tokens as a direct and clean representation of their deposited rugged tokens. The relationship between the deposited tokens and the minted $C_a$ tokens is given by:

\[
C_a = C_r
\]

These $C_a$ tokens function as receipts, representing the ownership and claim on the underlying rugged tokens. The $C_a$ tokens are fungible and can be freely used within the ecosystem, whether for trading, holding, or as collateral in other DeFi applications. The presence of $C_a$ tokens within the market serves as a signal that the user has deposited rugged tokens into the protocol, effectively transforming $C_r$ into a new, market-driven asset.

The market value of $C_a$ tokens is influenced by several factors, including:

1. \textbf{Market Supply and Demand}: The value of $C_a$ tokens, $P_{C_a}$, is determined by the market, where supply and demand dynamics play a crucial role:

\[
P_{C_a} = f(S_{C_a}, D_{C_a})
\]

As the demand for $C_a$ tokens increases relative to their supply, the market value $P_{C_a}$ will rise, and vice versa.

2. \textbf{Underlying Value of Rugged Tokens}: The value of $C_a$ tokens is also indirectly linked to the perceived value of the underlying rugged tokens $C_r$. If the market expects that the rugged tokens $C_r$ might regain value in the future, the value of $C_a$ could increase accordingly:

\[
P_{C_a} \propto E(P_{C_r})
\]

where $E(P_{C_r})$ is the expected future value of the rugged tokens.

3. \textbf{Liquidity and Utility}: The utility of $C_a$ tokens within the ecosystem—such as their use as collateral, in trading pairs, or in yield farming—contributes to their liquidity and market value. The more integrated and useful $C_a$ tokens are within the DeFi space, the higher their market value:

\[
P_{C_a} = g(U_{C_a}, L_{C_a})
\]

The protocol does not enforce any actions related to burning or altering the supply of $C_a$ tokens. Instead, the focus is on providing a straightforward receipt mechanism that empowers users to manage their assets independently within the ecosystem. This means that $C_a$ tokens remain purely market-driven assets, with their value being dictated by market conditions and the broader DeFi ecosystem. 

Users may choose to hold $C_a$ tokens, anticipating a rise in value due to market demand or the recovery of the underlying rugged tokens. Alternatively, they might trade or leverage $C_a$ in other DeFi applications, thereby influencing its liquidity and market presence.

This design choice simplifies the user experience while maintaining the integrity and transparency of the protocol. The absence of enforced mechanisms like burning or minting ensures that the value of $C_a$ tokens is wholly determined by the market, reflecting real-time supply and demand conditions within the ecosystem.

%%%%%%%%%%%%%%%%%%%%%%%

\section{Opportunities for New Ecosystems}
The Rugsafe protocol's structure introduces significant opportunities for developers and communities associated with rugged tokens $C_r$. Each rugged token $C_r$ represents an ecosystem that anticipated a specific offering or vision to materialize within the market. The issuance of $C_a$ tokens as a clean, fungible representation of $C_r$ provides these communities with a renewed opportunity to realize their original goals.

By depositing rugged tokens $C_r$ into a vault $V_c$, the protocol mints $C_a$ tokens in a 1:1 ratio:

\[
C_a = C_r
\]

These $C_a$ tokens, backed by provable deposits of the original underlying rugged tokens, allow communities to reintroduce their vision to the market under new terms. The $C_a$ tokens can be traded, utilized as collateral, or integrated into new decentralized finance (DeFi) applications, thereby reviving the ecosystem that was initially disrupted.

The presence of $C_a$ tokens in the market not only provides a second chance for these communities but also creates new financial products and services tailored to support the specific needs of rug pull victims. Developers can leverage this mechanism to build applications that cater to these communities, effectively transforming $C_r$ from a symbol of loss into a foundation for new economic activity.

Quantifying the impact, if $N$ is the total number of rugged tokens and $M$ represents the market value of each token, the introduction of $C_a$ creates a new market potential:

\[
\text{Market Potential} = \sum_{i=1}^{N} M_i \cdot C_{a,i}
\]

This market potential signifies the total value that can be recaptured or reintroduced into the ecosystem through $C_a$ tokens, providing a substantial incentive for developers and communities to participate in the Rugsafe protocol. The resulting ecosystem could not only restore lost value but also innovate new financial instruments and services, revitalizing the original vision of the rugged projects.

%%%%%%%%%%%%%%%%%%%%%%%%%%%

\subsection{User Commitment and Signal to the Ecosystem}

Depositing $C_r$ into a vault $V_c$ and receiving $C_a$ tokens signals the user's belief that $C_r$ is a rugged token. This action converts the rugged tokens into a new, clean representation within the ecosystem. To demonstrate a stronger commitment to the protocol, users can choose to burn their $C_a$ tokens, effectively removing them from circulation. This act of burning is a significant signal within the ecosystem.

The burning process can be represented as:

\[
\text{Signal} = \texttt{Burn}(C_a)
\]

Mathematically, the burning mechanism reduces the circulating supply of $C_a$ tokens, which can be expressed as:

\[
C_a' = C_a - B_{C_a}
\]

where \(C_a'\) is the circulating supply of $C_a$ after the burn, and \(B_{C_a}\) is the quantity of $C_a$ tokens burned.

This reduction in supply impacts the overall market dynamics and is reflected in the inverse relationship between the value of the Rugsafe token $R$ and the volume of rugged tokens $C_r$:

\[
R_\text{supply}  \propto \frac{1}{\log\left(\sum C_r \right)}
\]

The act of burning $C_a$ tokens strengthens this relationship by increasing the scarcity of $C_a$ and consequently affecting the value of $R$. The more $C_a$ is burned, the greater the signal of commitment to the protocol, potentially leading to a stronger inverse correlation between the value of $R$ and the total rugged tokens $C_r$.

\[
\frac{\partial R}{\partial C_a} < 0
\]

This partial derivative indicates that as the circulating supply of $C_a$ decreases (through burning), the value of $R$ is expected to increase, further solidifying the protocol's stability and the commitment of its participants.

%% could be problematic

%%%%%%%%%%%%%%%%%%%%%%%%%%%%%%%%%%%

\section{Rugsafe Vaults as Oracles}
Rugsafe vaults act as decentralized oracles, providing reliable data on the existence and volume of rugged tokens. This oracle functionality enables the protocol to adjust the supply of $R$ accurately and in real-time, based on the real-world data collected from the vaults.

%%%%%%%%%%%%%%%%%%%%%%%%%%%%%

%%%%%%%%%%%%%%%%%%%%%%%%

\section{Whale Penalty Mechanism}
To prevent large holders (whales) from gaming the system, the penalty for withdrawing rugged tokens $C_r$ scales with the amount of $C_a$ held. This mechanism is inspired by quadratic voting schemes, where the penalty increases progressively as the holder's balance of $C_a$ tokens grows, ensuring fairness and preventing manipulation. the penalty $P(C_a)$ for withdrawing rugged tokens is mathematically represented as:

\[
P(C_a) \propto \left(H_{C_a}\right)^\lambda
\]

Where $\lambda > 1$ is a scaling factor that ensures the penalty increases non-linearly with the amount of $C_a$ held.

\subsubsection{Impact on regular users vs. whales}

For regular users, who typically hold smaller amounts of $C_a$, the penalty remains relatively low, as the function’s exponent $\lambda$ ensures that the penalty grows slowly with their holdings:

\[
P_{\text{regular}}(C_a) = k \cdot \left(H_{C_a}\right)^\lambda \quad \text{with} \quad \lambda > 1 \quad \text{and} \quad H_{C_a} \text{ small}
\]

Where $k$ is a proportionality constant. for these users, the impact of the penalty is minimal, allowing them to interact with the protocol without significant costs, thus promoting fairness and inclusivity.

For whales, however, the penalty increases significantly due to their large holdings of $C_a$. as the amount of $C_a$ held by a whale increases, the penalty grows more steeply:

\[
P_{\text{whale}}(C_a) = k \cdot \left(H_{C_a}\right)^\lambda \quad \text{with} \quad \lambda > 1 \quad \text{and} \quad H_{C_a} \text{ large}
\]

This non-linear increase in penalty serves two purposes. first, it deters whales from attempting to manipulate the system by accumulating and withdrawing large amounts of $C_r$. Second, it ensures that whales, who may also be regular users of the protocol, are subject to the same rules but with penalties that reflect their larger influence on the system.

\subsubsection{Equitable system design}

\begin{figure}[h]
\centering
\includegraphics[width=\textwidth]{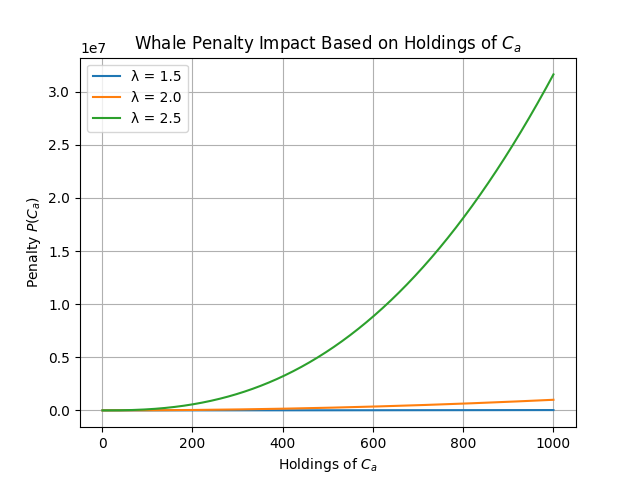}
\caption{the impact of the whale penalty mechanism based on varying scaling factors $\lambda$. as $\lambda$ increases, the penalty for holding larger amounts of $C_a$ grows exponentially, particularly affecting whales with significant holdings.}
\label{fig:whale_penalty}
\end{figure}

By scaling the penalty non-linearly with the amount of $C_a$ held, as in figure \ref{fig:whale_penalty}, the system remains equitable for all users, regardless of their holdings. Regular users are protected from disproportionate penalties, while whales, whose actions could have a more significant impact on the system, face penalties commensurate with their holdings. This design discourages potential manipulation by large holders and promotes a more balanced and fair ecosystem for all participants.

\subsubsection{Cumulative penalty mechanism}

\begin{figure}[h]
\centering
\includegraphics[width=\textwidth]{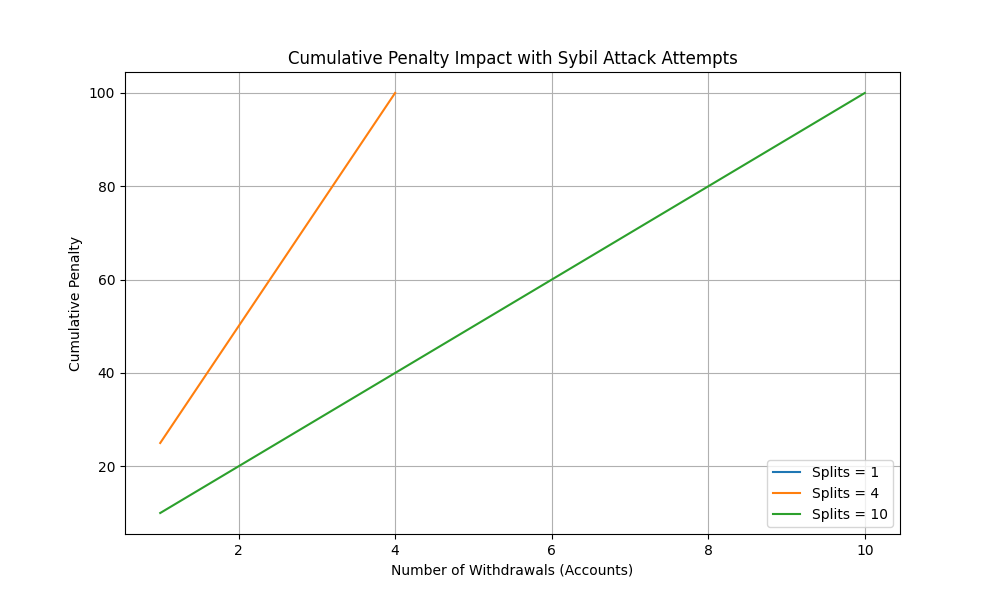}
\caption{the impact of the cumulative penalty mechanism on sybil attack attempts. the more accounts a whale creates to split their holdings, the higher the cumulative penalty becomes, discouraging such behavior.}
\label{fig:cumulative_penalty}
\end{figure}

To further enhance the protocol's defense against potential sybil attacks, a cumulative penalty mechanism has been introduced. This mechanism accumulates penalties progressively with each withdrawal attempt, particularly when a whale attempts to split their holdings across multiple accounts to evade higher penalties.

In this approach, the cumulative penalty is applied on a per-account basis but is aggregated across all related accounts. As illustrated in figure \ref{fig:cumulative_penalty}, the more accounts a whale creates to distribute their holdings, the more significant the overall penalty becomes. This cumulative effect discourages sybil attacks by ensuring that the total penalty incurred across all accounts exceeds what would have been paid if the whale had used a single account.

\[
P_{\text{cumulative}}(C_a) = P(C_a) \times \text{number of withdrawals}
\]

This means that while the whale may try to avoid the higher penalties by splitting their holdings across multiple accounts, the total cumulative penalty across those accounts will still be greater than the penalty they would have faced had they withdrawn the entire amount in a single transaction.

This can be seen in the plot below, where splitting across more accounts results in steeper cumulative penalties. the yellow and purple lines represent penalties for whales trying to split their holdings across 4 and 10 accounts, respectively.

%%%%%%

\subsubsection{Combining both mechanisms}

By combining the original non-linear penalty mechanism (which targets large holdings) with the cumulative penalty mechanism (which penalizes sybil attack attempts), the protocol ensures that whales are doubly disincentivized from trying to game the system. 

Whales face increasing penalties for holding large amounts of $C_a$. They also face additional cumulative penalties if they attempt to circumvent the system through sybil attacks by distributing their holdings across multiple accounts.

To formally prove that a whale attempting to split their holdings across multiple accounts incurs a higher penalty than withdrawing the entire amount in one transaction, consider the following:

\subsubsection{One-time withdrawal}

For a whale holding $H_{total}$ tokens, the penalty for a single withdrawal is:

\[
P_{\text{one-time}} = H_{total} \times \gamma
\]

Where $\gamma$ is the penalty rate (e.g., 10\%).

\subsubsection{Sybil withdrawal}

If the whale splits $H_{total}$ into $n$ accounts, the penalty for each account is given by:

\[
P_{\text{split}} = \left(\frac{H_{total}}{n}\right) \times \gamma
\]

The cumulative penalty is the sum of penalties across all $n$ accounts:

\[
P_{\text{cumulative}} = \sum_{i=1}^{n} \left( \frac{H_{total}}{n} \times \left( \gamma + \Delta \gamma \times i \right) \right)
\]

Where $\Delta \gamma$ is the incremental penalty applied to each subsequent withdrawal (i.e., cumulative penalty mechanism).

Simplifying this expression:

\[
P_{\text{cumulative}} = H_{total} \times \gamma + \frac{H_{total} \times \Delta \gamma \times n}{2}
\]

This expression shows that the cumulative penalty grows with both the number of splits $n$ and the incremental penalty $\Delta \gamma$, resulting in a total penalty that is higher than the one-time withdrawal penalty:

\[
P_{\text{cumulative}} > P_{\text{one-time}}
\]

Therefore, attempting to game the system through sybil attacks (splitting tokens across multiple accounts) leads to a higher overall penalty, as the cumulative penalty mechanism introduces increasing costs with each additional withdrawal.

This combined system ensures the integrity of the protocol by preventing manipulative behaviors and promoting fairness for all participants.

%%%%%%%%%%%%%%%%%%%%%%%%%%%%%%%%%%%%%

\section{Rugsafe Token Emissions and Parameterized Rewards Mechanism}

The Rugsafe protocol introduces a mechanism that incentivizes users to deposit rugged tokens $C_r$ into vaults $V_c$ by providing rewards in the form of Rugsafe tokens $R$. These vaults, serving as oracles, verify the deposits of specific rugged tokens and trigger emissions of Rugsafe tokens on the main chain.

In addition to receiving $C_a$, the user is rewarded with an airdrop of Rugsafe tokens $R$ on the main chain. The reward amount, $R_{\omega}$, for depositing rugged tokens is parameterized at the time of vault creation and is expressed as a function of the value of the deposit:

\[
R_{\omega} = f(C_r, \omega)
\]

where $\omega$ is a parameter set during vault creation that determines the emission rate tied to the value of the deposited tokens.

Furthermore, the protocol introduces a greater incentive for users who choose to burn their $C_a$ tokens, thereby signaling that they do not intend to redeem their underlying rugged tokens. The reward amount for burning, $R_{\texttt{burn}}$, is also parameterized at vault creation and is given by:

\[
R_{\texttt{burn}} = g(C_a, \theta) \quad \text{with} \quad g(C_a, \theta) > f(C_r, \omega)
\]

where $\theta$ is a parameter that sets a higher emission rate for burning $C_a$, indicating a permanent withdrawal of the rugged tokens from circulation.

Both the deposit and burn reward rates are flexible and can be tailored to the specific needs of the vault or community at the time of creation. This parameterization allows for customization of the incentives to align with the objectives of the protocol and the specific rugged token ecosystem.

This mechanism ensures that users are rewarded for contributing to the stability and security of the Rugsafe ecosystem. By depositing $C_r$ into vaults and potentially burning $C_a$, users receive Rugsafe tokens, fostering further participation and engagement in the protocol. The difference in reward levels between depositing and burning reflects the varying degrees of commitment to the ecosystem, with greater rewards for those who choose to fully transition away from their rugged tokens.

%%%%%%%%%%%%%%%%%%%%%%%%%%%%%%%%%%%%%%%%%

\section{Rugproof Mechanism}

The Rugsafe chain introduces a novel mechanism that allows users to mint new assets directly on the chain, with the assurance that these assets are rugproof by default. This mechanism is designed to protect users from potential rug pulls by leveraging a bonding system and a community-driven challenge process.

\subsubsection{Token Issuance and Bonding Mechanism}

At the time of token issuance, the issuer is required to put up a nontrivial percentage \(x\%\) of the total issued tokens as a bond:

\[
\beta_{\text{issuer}} = x\% \times T_I
\]

This bond acts as collateral, ensuring that the issuer has a stake in the integrity of the project. The bond is held in escrow and serves as a safeguard against potential rug pulls.

\subsubsection{Claim and Challenge Process}

At any point in time, any user can submit a claim against the token issuer, asserting that the project is being rugged. To submit a claim, the user must also put up a bond, \(\beta_{\text{claim}}\), as a guarantee of their claim:

\[
\beta_{\text{claim}} = y\% \times T_I
\]

where \(y\%\) is a parameter that may vary depending on the perceived risk or the specifics of the token.

Upon submitting a claim, a challenge period begins, during which other users can vote on whether the project is indeed rugging. Each vote requires a deposit, \(\delta_{\text{vote}}\), which can be chosen by the voter, though it may be smaller than the claim bond:

\[
\delta_{\text{vote}} \geq z
\]

where \(z\) is a minimum deposit amount set by the protocol.

\subsubsection{Outcome of the Challenge Period}

At the conclusion of the challenge period, the following outcomes are possible:

1. \textbf{Project Found to Be Rugging}: If the project is found to be rugging, the issuer's bond is slashed by a parameterized percentage, \(\alpha\%\), and distributed among all token holders, excluding the issuer:

\[
\beta_{\text{slashed}} = \alpha\% \times \beta_{\text{issuer}}
\]

The user who successfully claimed the rug pull receives a larger portion of the slashed bond, with the remaining amount distributed among the users who supported the claim.

2. \textbf{Claim Found to Be Fraudulent}: If the claim is found to be fraudulent, the claim bond is slashed by a parameterized percentage, \(\gamma\%\), and distributed to those who voted against the claim:

\[
\beta_{\text{slashed}_{\text{claim}}} = \gamma\% \times \beta_{\text{claim}}
\]

This process discourages false claims and encourages honest participation in the challenge process.

\subsubsection{Incentives and Protection}

This mechanism incentivizes users to carefully consider their actions when minting new assets and making claims. The initial bond posted by the issuer serves as a deterrent against rug pulls, while the challenge process allows the community to actively monitor and vote on the integrity of the project. By requiring bonds for claims and votes, the protocol ensures that participants have a stake in the outcome, reducing the likelihood of frivolous claims or votes.

By implementing this mechanism, the Rugsafe chain offers a robust, community-driven solution to protect users from potential rug pulls before they occur. This not only enhances the security of the ecosystem but also empowers users to take an active role in maintaining the integrity of the projects they support.

%%%%%%%%%%%%%%%%%%%%%%%%%%%%%%%%%%%%%%%

%%%%%%%%%%%%%%%%%%%%%%%%%%%%%
%%%%%%%%%%%%%%%%%%%%%%%%%%%%%
%%%%%%%%%%%%%%%%%%%%%%%%%%%%%
%%%%%%%%%%%%%%%%%%%%%%%%%%%%%
%%%%%%%%%%%%%%%%%%%%%%%%%%%%%
%%%%%%%%%%%%%%%%%%%%%%%%%%%%%

% DEFI PRIMITIVES
\section{DeFi for Rugged Tokens}

The \textbf{DeFi Suite} introduces advanced financial instruments designed to empower users holding rugged tokens, transforming these otherwise illiquid assets into valuable opportunities for growth and profit. Among these instruments, \textbf{perpetual futures} play a pivotal role in creating new market dynamics for token holders, providing unique methods for hedging, speculation, and liquidity.

%%%%%%%%%%%%%%%
%%%%%%%%%%%%%%%%%%%%%%%%%%%%%
%%%%%%%%%%%%%%%%%%%%%%%%%%%%% DEX

\subsection{Decentralized Exchange}

The Decentralized Exchange (DEX) is a critical infrastructure component of the protocol, providing a platform for the trading and liquidity provision of rugged tokens, anticoins, and other assets within the ecosystem. The DEX facilitates seamless swaps between these assets, ensuring that even illiquid rugged tokens can find liquidity within the Rugsafe ecosystem.

By enabling users to trade and provide liquidity, the DEX creates new financial opportunities for rug pull victims, transforming otherwise stagnant assets into dynamic trading vehicles. Key features of the DEX include liquidity provision, trading, and access to new financial products. Users can provide liquidity to trading pairs, such as \textit{USDC/aTokenA} or \textit{USDC/TokenA}, earning fees on trades and helping maintain liquidity for rugged tokens and anticoins. The trading functionality allows users to swap rugged tokens for anticoins or other stable assets, giving them flexibility in managing their positions and hedging against further declines in rugged token prices. Furthermore, the DEX serves as a hub for innovative financial products tied to rugged tokens, including synthetic assets, derivatives, and collateralized loans, enabling a vibrant market for previously illiquid tokens.

The Rugsafe DEX ensures that the protocol remains liquid and adaptable, offering users various ways to interact with and extract value from their rugged tokens and anticoins.

\subsubsection{Enforcement of the Logarithmic Inverse Peg}

One of the DEX’s most important functions is the enforcement of the logarithmic inverse peg, which governs the relationship between rugged tokens and anticoins. The inverse peg ensures that as the value of a rugged token decreases, the value of its corresponding anticoin increases, providing users with a natural hedge against their rugged token holdings.

The DEX plays a central role in maintaining this peg by facilitating swaps between rugged tokens and anticoins, with the price of anticoins adjusting in real time based on market conditions. The automated market maker (AMM) within the DEX continuously updates the price relationship between rugged tokens ($C_r$) and anticoins ($C_a$), enforcing the inverse peg:

\[
C_{a}(t) = \log\left(\frac{C_{r}(0)}{C_{r}(t)}\right)
\]

Where \( C_{a}(t) \) is the value of the anticoin at time \( t \), and \( C_{r}(t) \) is the real-time price of the rugged token. By automatically adjusting the price of anticoins as rugged token prices fluctuate, the DEX ensures that the inverse peg remains intact, providing users with a reliable way to hedge against further declines in rugged token value.

%%%%%%%%%%%%%%%%%%%%%%%%%%%%%
%%%%%%%%%%%%%%%%%%%%%%%%%%%%%
%%%%%%%%%%%%%%%%%%%%%%%%%%%%%
%%%%%%%%%%%%%%%%%%%%%%%%%%%%%
%%%%%%%%%%%%%%%%%%%%%%%%%%%%%
%%%%%%%%%%%%%%%%%%%%%%%%%%%%%
%%%%%%%%%%%%%%%%%%%%%%%%%%%%%

\subsection{Perpetual Futures}

In a typical perpetual contract, users can open \textit{long} or \textit{short} positions on an asset without an expiration date, profiting from price movements in either direction. The Rugsafe protocol extends this concept to \textit{rugged tokens} through its vault system, allowing holders of rugged tokens to open perpetual positions using \textbf{anticoins} as collateral. This mechanism enables users to leverage their token holdings to speculate on the future price movements of rugged tokens, even if the tokens themselves are considered worthless.

Holders of \textbf{anticoins} can open perpetual positions by locking their collateral in the \textbf{Rugsafe Perpetuals System}. Users can profit from the token's price volatility by choosing to either long or short rugged tokens. This system offers an innovative way for rug pull victims to continue interacting with their assets, providing new avenues for financial recovery.

\[
P_{\text{perpetual}}(t) = f\left( C_a, \ell, \text{pos} \right)
\]

where \( P_{\text{perpetual}}(t) \) is the value of the perpetual position over time, \( C_a \) is the anticoin collateral locked by the user, \( \ell \) represents the leverage chosen by the trader, and \( \text{pos} \) represents the direction of the position (either \textit{long} or \textit{short}).

\subsubsection{Asymmetric Funding Rates}

A critical component of perpetual contracts is the \textbf{funding rate} mechanism, which ensures market balance between long and short positions. In the Rugsafe protocol, asymmetric funding rates are employed to provide additional incentives for liquidity providers (LPs), especially when dealing with rugged tokens that lack intrinsic value. 

The asymmetric funding rate ensures that one side of the market—either the long or short side—bears a higher cost based on the market's position imbalance. This imbalance, particularly when rugged tokens are involved, often leans toward the short side, as participants may be more inclined to bet against the token’s value. To balance this out, liquidity providers who take the less popular side (e.g., the long side for a heavily shorted rugged token) are rewarded with higher funding rate payments.

This creates a scenario where LPs are incentivized to provide liquidity, even for tokens with little to no intrinsic value, because they can earn higher returns from funding rate differentials. For example, in a market where the majority of participants are shorting a rugged token, the long side receives higher funding rate payments, encouraging liquidity provision.

\[
F_{\text{rate}}^{\text{long}} \neq F_{\text{rate}}^{\text{short}}
\]

The funding rates \( F_{\text{rate}}^{\text{long}} \) and \( F_{\text{rate}}^{\text{short}} \) are dynamically adjusted based on market conditions. If the imbalance between long and short positions is extreme, the funding rate paid by shorts will be significantly higher than the rate received by longs, and vice versa, to incentivize participation on the less popular side.

%%%%%%% FUNDING RATE
%%%%%%%%%%%%%%%%%%

\subsubsection{Funding Rate Calculation}

Funding rates are calculated and paid at regular intervals, ensuring that perpetual contract holders are either paying or receiving funding rates based on the market’s position dynamics. The calculation is driven by the ratio of long and short positions, with the side having fewer participants (e.g., long positions in a rugged token) receiving favorable funding rates. The protocol automatically adjusts these rates to maintain equilibrium and continually incentivize liquidity providers.

\[
F_{\text{rate}} = \alpha \times \left( 1 - \frac{N_{\text{short}}}{N_{\text{long}} + N_{\text{short}}} \right)
\]

where \( F_{\text{rate}} \) represents the funding rate, \( N_{\text{long}} \) and \( N_{\text{short}} \) represent the number of long and short positions, respectively, and \( \alpha \) is a constant that determines the base funding rate.

This formula ensures that the side with fewer positions receives a higher funding rate, incentivizing liquidity provision. The funding rate is dynamic, adjusting based on the real-time market conditions to stabilize long and short participation. Additionally, liquidity depth \( L_{\text{pool}} \) is taken into consideration to avoid large funding rate spikes in low liquidity scenarios. The overall funding rate is scaled by liquidity depth as follows:

\[
F_{\text{rate,final}} = F_{\text{rate}} \times \left( 1 + \frac{L_{\text{min}}}{L_{\text{pool}}} \right)
\]

where \( L_{\text{min}} \) is a minimum liquidity threshold, and \( L_{\text{pool}} \) represents the current liquidity depth in the pool. This ensures that funding rates remain manageable even in scenarios of low liquidity, preventing extreme funding spikes. This asymmetric funding model encourages active participation from liquidity providers, even in scenarios where the rugged token has little to no value, as they can profit from the differences in funding rates between long and short positions.

\subsubsection{Liquidation Mechanism and Protocol-Owned AMM}

To manage risk and ensure market stability, the protocol employs a hybrid liquidation mechanism. Positions that become under-collateralized are first flagged for liquidation. Liquidators in the network are given an opportunity to execute these liquidations and earn a profit by doing so. If a liquidator fails to execute the liquidation within a predefined timeframe, the protocol initiates automated liquidation to prevent prolonged under-collateralization. Liquidation events trigger interactions with the protocol-owned AMM, where rugged tokens or anticoins are swapped for liquid assets to pay liquidators or to settle the debt. The protocol-owned AMM ensures that liquidators and liquidity providers are compensated fairly, even when rugged tokens are illiquid. The AMM acts as a buffer, converting illiquid rugged tokens into more liquid assets to maintain stability in the perpetual markets.

%%%%%%%%%%%%%%%%%%%%%%%%%%%%%
%%%%%%%%%%%%%%%%%%%%%%%%%%%%%
%%%%%%%%%%%%%%%%%%%%%%%%%%%%%
%%%%%%%%%%%%%%%%%%%%%%%%%%%%%
%%%%%%%%%%%%%%%%%%%%%%%%%%%%%
%%%%%%%%%%%%%%%%%%%%%%%%%%%%%
%%%%%%%%%%%%%%%%%%%%%%%%%%%%%

%%%%%%%%%%%%%%%%%%%%%%%%%%%%% Rug Detector

\section{Rug Detection Mechanisms}

The \textbf{Rug Detection Mechanism} is a core component of the Rugsafe protocol, designed to identify and respond to potential rug pulls in real-time. By monitoring liquidity pools and detecting early signs of liquidity removal, the system aims to protect users and mitigate the impact of rug pulls through strategic actions, such as frontrunning, sandwiching, or backrunning. The rug detection system operates under the assumption that liquidity is not removed instantly, but rather over a short period, providing a critical window for intervention. Our on-chain detectors combine economic signals with static-analysis ideas first explored in \cite{lin2024} to flag risky smart-contracts before liquidity vanishes. We also ingest live transaction graphs, using fusion techniques similar to \cite{wu2025}, to catch stealth pulls that slip past pure static analysis.

\subsection{Liquidity Pool Monitoring}

The primary method of rug detection is through the continuous monitoring of liquidity pools, where most rug pulls begin. The protocol tracks key metrics in liquidity pools, including liquidity volumes, token price movements, and the ratio between paired assets. A significant and sudden reduction in liquidity is often an early indicator of an impending rug pull.

\[
\Delta L_{\text{pool}} = L_{\text{pool}}(t) - L_{\text{pool}}(t-1)
\]

where \( \Delta L_{\text{pool}} \) represents the change in liquidity over time. A large negative value of \( \Delta L_{\text{pool}} \) indicates that liquidity is being withdrawn from the pool. The protocol automatically detects these changes in liquidity and responds by initiating strategic actions to mitigate potential losses for users.

\subsection{Front-running Liquidity Removal}

In cases where liquidity is being gradually removed, the protocol can attempt to \textbf{front-run} the rug pull by placing a transaction before the liquidity is fully withdrawn. By doing so, the protocol can secure better prices for affected users before the liquidity is fully drained, protecting their positions. Front-running is particularly effective in liquidity pools where large liquidity withdrawals are observable on-chain before they are finalized. The protocol submits a buy or sell order before the liquidity provider completes their transaction, enabling Rugsafe users to escape before the token price collapses.

\subsection{Sandwich Attacks on Liquidity Withdrawal}

An alternative strategy is the \textbf{sandwich attack}, where the protocol places transactions both before and after the liquidity withdrawal. In this scenario, the protocol attempts to maximize the price slippage created by the rug pull. The sandwich consists of three steps:

\begin{enumerate}
    \item The protocol detects the liquidity removal transaction and submits a \textit{pre-transaction} that takes advantage of the available liquidity.
    \item The liquidity removal transaction executes, causing price slippage in the pool.
    \item The protocol submits a \textit{post-transaction} that profits from the new price created by the liquidity removal.
\end{enumerate}

This strategy allows the protocol to capitalize on the price volatility caused by liquidity withdrawals, potentially earning profits for the Rugsafe treasury or users participating in the detection process.

\subsection{Back-running Liquidity Removal}

\textbf{Back-running} is another key strategy used by the protocol. In this case, the protocol submits a transaction immediately after liquidity is removed, taking advantage of the price changes that occur as a result of the rug pull. Back-running allows the protocol to capture residual value left in the market after the liquidity has been drained. While this strategy may not fully prevent losses for users, it can reduce the overall impact by allowing the protocol to act quickly in the moments following liquidity removal.

\subsection{Detection Window and Response Time}

To ensure the effectiveness of rug detection, the protocol implements a tight window for detection and response. Rug pulls typically do not happen instantaneously, especially in more established tokens with active trading. This gradual removal of liquidity provides a window of opportunity, where the protocol can detect and react, leveraging blockchain transaction ordering mechanisms. The detection system leverages on-chain data and off-chain tools to monitor liquidity pools in real-time, ensuring that potential rug pulls are identified as quickly as possible. The system’s response time is governed by the block time of the underlying blockchain, and it can initiate automated transactions before the liquidity is fully withdrawn.

\subsection{Frontrunning Prevention and Countermeasures}

The protocol also incorporates measures to prevent malicious actors from using front-running techniques to game the system. By setting transaction fees and gas limits dynamically, the protocol can ensure that its own protection mechanisms remain operational even in the presence of high gas fee spikes or congested networks. Additionally, the protocol deploys strategies to minimize the risk of other bots or malicious actors frontrunning its transactions, ensuring that Rugsafe retains priority when submitting critical protection transactions.

\subsection{Other Rug Detection Metrics}

In addition to monitoring liquidity withdrawals, the Rugsafe detection mechanism incorporates a variety of other metrics that may indicate a potential rug pull, including:

\begin{itemize}
    \item \textbf{Token Minting}: A sudden increase in token supply or minting events could indicate a malicious attempt to flood the market before removing liquidity.
    \item \textbf{Wallet Activity}: Monitoring large transfers from developer or creator wallets can provide early warnings of suspicious behavior.
    \item \textbf{Unusual Trading Volumes}: A spike in trading volumes without corresponding increases in liquidity could be a signal of impending manipulation.
\end{itemize}

These additional metrics provide a more comprehensive view of the market, ensuring that the protocol can detect a wide range of rug pull scenarios and protect users accordingly.

\subsubsection{Protocol Governance and Community Involvement}

To further enhance the rug detection system, the Rugsafe community can participate in governance decisions that influence how the detection mechanism operates. For example, the community can vote on which liquidity pools to prioritize, adjust the detection sensitivity, and allocate protocol resources to further improve the detection infrastructure. This decentralized governance model ensures that the rug detection system evolves alongside the broader ecosystem, continuously adapting to new threats and opportunities in decentralized finance.

\subsection{Intent-Based Mitigation System}

To complement the real-time rug detection mechanisms, Rugsafe introduces an \textbf{intent-based mitigation system} that empowers users to proactively safeguard their assets against potential rug pulls. This system allows users to predefine conditions for exiting positions or swapping their rugged tokens before significant losses occur, providing an additional layer of automated protection.

\subsubsection{Submitting Intents}

Users can submit intents specifying conditions under which their positions should be liquidated or their rugged tokens swapped for more stable assets. For example, a user may submit an intent to exit their position if liquidity in the pool drops below a certain threshold or if the token price declines by a predefined percentage.

\[
I = f\left( P_{r}(t), L_{\text{pool}}(t), \theta \right)
\]

where \( I \) is the intent submitted by the user, \( P_{r}(t) \) is the real-time price of the rugged token, \( L_{\text{pool}}(t) \) is the liquidity level in the pool, and \( \theta \) represents the user-defined thresholds for triggering the intent.

\subsubsection{Solver Role in Executing Intents}

Once an intent is triggered, \textbf{solvers} within the Rugsafe network act on behalf of the user by executing the necessary transactions. Solvers are responsible for continuously monitoring on-chain conditions, ensuring that when a user's predefined conditions are met, the appropriate actions are taken to exit positions or swap tokens.

Solvers are incentivized by receiving a portion of the transaction fees or a small reward in exchange for their services, ensuring that user intents are executed swiftly and effectively.

\subsubsection{Mitigation as a Safeguard}

The intent-based mitigation system acts as a safeguard for users, providing automated risk management even when they are not actively monitoring the protocol. By setting up predefined exit strategies, users can protect themselves from the impact of rug pulls or sudden liquidity changes without needing to react in real-time. This system transforms rug pull prevention from a reactive to a proactive approach, ensuring that users' funds are safeguarded against market manipulation or malicious activities in decentralized finance.

%%%%%%%%%%%%%%%%%%%%%%%%%%%%%
%%%%%%%%%%%%%%%%%%%%%%%%%%%%%
%%%%%%%%%%%%%%%%%%%%%%%%%%%%%
%%%%%%%%%%%%%%%%%%%%%%%%%%%%%
%%%%%%%%%%%%%%%%%%%%%%%%%%%%%
%%%%%%%%%%%%%%%%%%%%%%%%%%%%%

%%%%%%%%%%%%%%%%%%%%%%%%%%%%%
%%%%%%%%%%%%%%%%%%%%%%%%%%%%%
%%%%%%%%%%%%%%%%%%%%%%%%%%%%%
%%%%%%%%%%%%%%%%%%%%%%%%%%%%%
%%%%%%%%%%%%%%%%%%%%%%%%%%%%%
%%%%%%%%%%%%%%%%%%%%%%%%%%%%%
%%%%%%%%%%%%%%%%%%%%%%%%%%%%%
%%%%%%%%%%%%%%%%%%%%%%%%%%%%% INSURANCE

\section{Decentralized Insurance Mechanism}

The Decentralized Insurance Mechanism (DIM) introduces a community-driven system for users to protect their holdings from potential risks associated with decentralized finance activities, such as smart contract failures, protocol exploits, and market manipulation. This insurance model leverages a dispute-resolution process involving bonds, claims, and challenge periods, similar to the \textbf{Rugproof Mechanism}, but tailored specifically for coverage and compensation in decentralized insurance events.

\subsection{Insurance Issuance and Bonding Mechanism}

At the time of insurance issuance, the insurer is required to put up a bond as collateral to ensure that they have a stake in the integrity of the coverage. The bond is designed to incentivize honest behavior from both the insurer and the claimant. The bond value, $\beta_{\text{insurer}}$, is a percentage of the total insured value, $I_v$, determined by the protocol:

\[
\beta_{\text{insurer}} = x\% \times I_v
\]

The bond ensures that the insurer will honor valid claims and act in good faith throughout the policy's lifespan. The bond is held in escrow during the policy's duration, and in the event of a dispute, it may be slashed or forfeited based on the community’s decision.

\subsection{Submitting a Claim and Claim Bonds}

If an insured event occurs (e.g., a smart contract exploit or market failure), the user can submit a claim to recover their losses. To initiate a claim, the user must also deposit a bond, $\beta_{\text{claim}}$, as a commitment to their claim's legitimacy. This claim bond serves as a deterrent for fraudulent or frivolous claims.

\[
\beta_{\text{claim}} = y\% \times I_v
\]

Once the claim is submitted, a challenge period begins. During this period, other participants in the protocol can review the claim and either approve it or dispute it. The challenge period duration, $\tau_{\text{challenge}}$, is parameterized based on the type of insurance policy, providing enough time for disputes to be raised if necessary.

\subsection{Challenge Period and Dispute Mechanism}

During the challenge period, any participant in the protocol can challenge the claim if they believe it is fraudulent or inaccurate. Challengers are required to deposit a dispute bond, $\beta_{\text{dispute}}$, to initiate the dispute process:

\[
\beta_{\text{dispute}} = z\% \times I_v
\]

If a dispute is raised, the protocol enters an escalation phase, where both the claimant and the challenger must present their cases. The dispute is then resolved through a community voting process, where users stake their tokens to vote on the validity of the claim. The voting period, $\tau_{\text{vote}}$, is designed to give ample time for the community to review the evidence and cast their votes.

\subsection{Outcome of the Dispute Process}

At the conclusion of the voting period, there are two possible outcomes:

\subsubsection{Claim Approved}
If the claim is found to be valid, the insurer is required to compensate the claimant according to the terms of the insurance policy. The insurer’s bond, $\beta_{\text{insurer}}$, may be partially or fully slashed, depending on the severity of the claim. The claimant receives the insured amount, $I_v$, and a portion of the insurer’s bond as compensation for their losses.

\[
\text{Compensation} = I_v + \alpha \cdot \beta_{\text{insurer}}
\]

Where $\alpha$ is a slashing factor that determines how much of the insurer’s bond is distributed to the claimant.

\subsubsection{Claim Rejected}
If the claim is found to be fraudulent or inaccurate, the claimant’s bond, $\beta_{\text{claim}}$, is slashed and distributed to the challenger and the community members who voted against the claim. This discourages false claims and ensures that the dispute process remains fair.

\[
\text{Penalty} = \gamma \cdot \beta_{\text{claim}}
\]

Where $\gamma$ is the slashing factor applied to the claimant's bond.

\subsection{Escalation and Final Resolution}

If either party is unsatisfied with the outcome of the dispute, they can escalate the case to a higher arbitration level, where additional community governance and smart contract verification may be invoked to reassess the claim. During escalation, both the claimant and challenger may be required to post additional bonds, further increasing their stake in the dispute’s outcome.

The escalation process ensures that malicious behavior is discouraged at all levels, as parties who act dishonestly risk losing even more collateral in the event of an unfavorable outcome.

\subsection{Incentives for Honest Behavior}

The decentralized insurance mechanism is designed to incentivize honest behavior from all participants. Insurers are encouraged to act in good faith by having their bonds at stake, while claimants and challengers are deterred from submitting or disputing false claims due to the risk of losing their bonds. Community members who participate in the voting process are rewarded for correctly identifying valid claims and challenging fraudulent ones.

By utilizing bonds, challenge periods, and community-driven voting, the Rugsafe decentralized insurance system provides a fair, transparent, and secure method for users to insure their assets against risks in decentralized finance.

%%%%%% Collective claims 

\subsection{Collective Claims and Bond Pooling}

In cases where multiple users are affected by the same rug pull event, they can join an existing claim to collectively seek compensation. When joining a claim, each user is required to post a bond, $\beta_{\text{join}}$, similar to the original claim bond. This bond serves as collateral, ensuring that all participants in the claim are incentivized to submit only valid and truthful claims.

\[
\beta_{\text{join}} = w\% \times I_v
\]

where $w\%$ is the bond percentage determined by the protocol for users joining an existing claim.

Each user who joins the claim agrees to be bound by the same outcome as the original claimant. If the collective claim is approved, all participants receive a share of the compensation fund, distributed proportionally based on their submitted loss amounts. However, if the collective claim is found to be fraudulent, the bonds of all participants who joined the claim are pooled together and subject to slashing.

\[
\text{Total Penalty} = \gamma \cdot \sum_{i=1}^{n} \beta_{\text{join}, i}
\]

where $n$ is the number of participants in the collective claim, and $\gamma$ is the slashing factor applied to the pooled bond amount.

This collective bond slashing mechanism ensures that all participants have a vested interest in the claim’s validity. Users are incentivized to verify the legitimacy of a claim before joining, as they risk losing their bond if the claim is later deemed fraudulent. Additionally, by pooling bonds and distributing penalties across multiple participants, the protocol strengthens the overall integrity of the insurance system.

By allowing users to join claims and pool their bonds, Rugsafe’s decentralized insurance mechanism provides an inclusive and scalable way to address losses from rug pulls, offering collective security and shared responsibility.

%%%%%%%%%%%%%%%%%%%%%%%%%%%%%
%%%%%%%%%%%%%%%%%%%%%%%%%%%%%
%%%%%%%%%%%%%%%%%%%%%%%%%%%%%
%%%%%%%%%%%%%%%%%%%%%%%%%%%%%
%%%%%%%%%%%%%%%%%%%%%%%%%%%%%
%%%%%%%%%%%%%%%%%%%%%%%%%%%%%
%%%%%%%%%%%%%%%%%%%%%%%%%%%%%

\section{Closing Remarks}
The prevalence of rug pulls represents one of the most pressing threats to the integrity and trust in the cryptocurrency ecosystem. Rugsafe is committed to addressing this challenge by leveraging innovative mechanisms to protect users from rug pulls, and enable new opportunities for rugged investors. By providing a secure framework for asset recovery and offering tools that prevent malicious activity, Rugsafe aims to eliminate one of the largest sources of financial loss in the industry. Rugsafe’s mission is to transform vulnerabilities into resilience, ensuring digital assets are protected and confidence in the industry is restored.

\end{document}